\documentstyle[12pt]{article}
\begin{document}

\begin{center}{\Large {\bf Quantum effects in classical chaos through intermittency}}\end{center} 

\vskip 0.5cm

\begin{center}{T. S. Nag  and P.Banerjee}\\
{Department of Physics, Presidency University\\
86/1 College Street, Kolkata - 700 073, India}\\
{E-mail:banprabir@gmail.com}\end{center}

The semi-quantal dynamics is applied to investigate the influence of quantum 
fluctuations on problems in classical chaos through intermittency involving
bifurcations. The results of the numerical calculations indicate that quantum 
effects enhance the tendency to chaos in both the problems of inverted 
pitchfork and saddle-node bifurcations discussed here.

\begin{flushleft}{PACS Nos.: 05.45.-a,05.45.Gg,05.45.Mt}\end{flushleft}

\newpage

\section{Introduction}

Intermittency is one of the well-established routes to chaos, particularly in problems of fluid
turbulence and in certain problems of reaction diffusion systems\cite{guck,pom,hu1,la}. 
The appearance of alternate regions of regular and chaotic phases with the relative durations of two regions determined by the parameters of the system is the hallmark of this route. The saddle-node and the inverted pitchfork bifurcations 
\cite{guck,pom,hu1,lah} form two important classes of bifurcations leading to intermittency. It is to be noted that while the saddle-node bifurcation is a typical codimension-1 bifurcation leading to type-1 intermittency, the inverted pitchfork bifurcation combines features of type-2 as well as type-3 intermittencies in the sense that an unstable branch of fixed points persists on the chaotic side of the bifurcation point. An exact renormalization group analysis of these bifurcations is also available in literature\cite{hu1,lah}.

Since quantum effects are ubiquitous, it is interesting to look into the effects of quantum fluctuations on intermittency. The introduction of quantum effects 
on a kicked rotor, that leads in the classical theory to the onset of chaos 
above a critical value of the parameter, results in the suppression of chaos, 
similar to Anderson's localisation in the presence of disorder\cite{haa,chi}. 
The quantum 
fluctuations have been shown in our work to enhance chaos in both the 
types of intermittency (saddle-node and inverted pitchfork) referred to earlier.
This enhancement is reflected in the three well-known features: (i) decrease of
average staying time in the laminar (or regular) phase, (ii) increase of 
Lyapunov exponent, (iii) increased rate of diffusion in energy space. 
The phase space plots will also enlighten us about the quantum effects on the
classical evolution.

The squeezed coherent state formalism
\cite{won,hu2,jac,zhan,pat1,pat2,pat3,zhang} has been adopted in presenting a 
semi-quantal analysis of the theory. A similar analysis, followed also in other 
cases, has been seen to lead to enhancement as well as suppression of chaos depending 
on the examples chosen\cite{hu2}. Starting with the Hamiltonian equations relevant to our
problem, we have obtained certain model difference equations that give rise to intermittencies
in the classical limit. This work is, we believe, the first attempt towards understanding the
effect of quantum fluctuations on problems involving intermittency.

Section II discusses briefly the theoretical background of the numerical calculations and Section III the numerical results. The last section concludes with the summary as well as the discussion of the results obtained.
  
\section{Formalism}
The squeezed state formalism\cite{won,hu2} stems from the time-dependent variational principle
\begin{eqnarray}
\Delta \int dt\langle \psi(t) \mid i\hbar {\frac\partial {\partial t}} - {\hat H}\mid \psi(t) \rangle = 0,
\label{eq1}
\end{eqnarray}
where $\psi(t)$ is chosen to be the squeezed coherent state represented by 
$\mid \psi(t)\rangle = D(\alpha) S(\gamma)\mid0\rangle$, $D(\alpha) = \exp (\alpha a^{\dagger} - \alpha ^{\star}a)$, $S(\gamma) = \exp [{\frac 1 2}(\gamma {a^{\dagger}}^2
- \gamma ^{\star}a^2)]$ and $\mid0\rangle$ represents the vacuum state.
The pairs of canonical variables given by ${\hat x} =\sqrt{\frac\hbar 2}(a^{\dagger} +a)$, $ {\hat p}=i\sqrt{\frac\hbar 2}(a^{\dagger}-a)$, $\Delta x^2=(x-\langle x\rangle)^2$, $\Delta p^2=(p-\langle p\rangle)^2$  are obtained as 
\begin{eqnarray}
x=\langle\psi\mid {\hat x}\mid\psi\rangle = \sqrt{\frac\hbar 2}(\alpha^{\star} 
+\alpha),\nonumber\\ p= \langle\psi\mid {\hat p}\mid\psi\rangle = i\sqrt{\frac\hbar 2}(\alpha^{\star} -\alpha),\nonumber\\
\Delta x^2=\langle\psi\mid({\hat x}- x)^2\mid\psi\rangle=\hbar \phi,\nonumber\\ 
\Delta p^2=\langle\psi\mid({\hat p} - p)^2\mid\psi\rangle
 =\hbar({\frac1 {4\phi}} + 4\pi^2 \phi),\nonumber\\
\phi={\frac 1 2}[\cosh\mid\gamma\mid+{\gamma \over \mid\gamma\mid}
\sinh\mid\gamma\mid]^2,\nonumber\\
\pi={\frac i 4} {(\gamma^{\star}-\gamma) \over \mid\gamma\mid}
{\sinh2\mid\gamma\mid  \over (\cosh\mid\gamma\mid + 
{\gamma \over \mid\gamma\mid} \sinh\mid\gamma\mid)^2}\nonumber
\end{eqnarray}
Clearly $(x,p)$ represent the centroids of the wavepacket in the coordinate and
the momentum space respectively while $\Delta x^2$ and $\Delta p^2$ give 
measures of the quantum fluctuations in the coordinate and the momentum space
respectively. They satisfy the canonical equations 
\begin{eqnarray}
{\dot x} = {\frac {\partial H} {\partial p}},
{\dot p} = -{\frac{\partial H} {\partial x}},\nonumber\\
\hbar {\dot \phi} = {\frac{\partial H} {\partial \pi}},
\hbar {\dot \pi} =-{\frac{\partial H} {\partial \phi}}\nonumber,
\end{eqnarray}
where the Hamiltonian function $H$ given by $\langle\psi\mid {\hat H}\mid\psi\rangle$ can be expressed as  
\begin{eqnarray}
H = {\frac {p^2} 2} + V(x) + \hbar({\frac 1 {8\phi}} + 2\pi^2 \phi)  
+\{\exp[{\frac{\hbar \phi} 2}({\partial\frac \partial x})^2] - 1\}V(x).
\label{eq2}
\end{eqnarray}

\subsection{CASE I (Inverted pitchfork bifurcation)}

The mapping giving rise to the inverted pitchfork bifurcation in two dimensions
corresponds to the periodically kicked Hamiltonian given by
\begin{eqnarray}
H = {\frac {p^2} 2} + {\frac {\Delta p^2} 2} - [({\frac {\epsilon x^2} 2} + 
{\frac{\beta x^4} 4}) + {\frac{\Delta x^2} 2}(\epsilon + 3\beta x^2) 
+6\beta {\frac{(\Delta x^2)^2} 8}]\sum_n\delta(t-nT),
\label{eq3}
\end{eqnarray}
$T$ being the time interval between two successive kicks.

Substituting for $\Delta x^2$ and $\Delta p^2$ from Eq.~(\ref{eq2}) we obtain 
\begin{eqnarray}
H = [{\frac {p^2} 2} + \hbar({\frac 1 {4\phi}} + 4\pi^2 \phi)]  
- [{\frac{\epsilon x^2} 2} + 
{\frac{\hbar \phi} 2}(\epsilon + 3\beta x^2) 
+{\frac{3\hbar ^2\phi ^2\beta} 4}]\sum_n\delta(t-nT),
\label{eq4}
\end{eqnarray}

The Hamiltonian in Eq.~(\ref{eq4}) will yield the following difference equations after 
utilising Eq.~(\ref{eq2}) and going through algebraic manipulations:
\begin{eqnarray}
X_{n+1}=X_n + P_n,
\label{eq5}
\end{eqnarray}
\begin{eqnarray}
P_{n+1}=P_n + T[(\epsilon X_n + \beta X_n^3) + 
{\frac{3\beta \hbar X_n\Phi _n T} 2}],
\label{eq6}
\end{eqnarray}
\begin{eqnarray}
\Phi _{n+1} = \Phi _nF_n, F_n = {\frac 1 {\Phi _n^2}} + (\Pi _n+1)^2
\label{eq7}
\end{eqnarray}
and
\begin{eqnarray}
\Pi _{n+1} = [\Pi _n + \Pi _n^2 + {\frac 1 {\Phi _n^2}}]{\frac 1 F_n} + 
2T\Delta _{n+1},
\label{eq8}
\end{eqnarray}
where
$\Delta _{n+1} = {\frac 1 2}[(\epsilon + 3\beta X_n^2) + {\frac{3T\beta \hbar 
\Phi _n} 2}]$, $X_n = x_n, P_n = p_nT, \Phi _n = {\frac{2\phi _n} T},
\Pi _n = 2T\pi _n$.

The Eqs.~(\ref{eq5}) and ~(\ref{eq6}) give, in the limit $\hbar \rightarrow 0$, the classical
inverted pitchfork bifurcation. This is characterised by the fact that there
exist three fixed points $X = 0, \pm \sqrt(-{\frac \epsilon \beta}), P = 0$
for $\epsilon < 0, \beta > 0$, the first one being stable, the other two being
unstable, going over to only one fixed point (unstable) at $X=0$ for $\epsilon
> 0, \beta >0$. The situation pertaining to  
$\epsilon > 0, \beta < 0$ creates two stable fixed points at 
$X = 0, \pm \sqrt(-{\frac \epsilon \beta}), P = 0$ and one unstable fixed point
at $X = 0, P = 0$, while for $\epsilon < 0, \beta < 0$ the latter one becomes 
stable and the earlier ones vanish. The first case ($\beta > 0$) corresponds to
the subcritical and the second one ($\beta < 0$) to supercritical bifurcations
respectively in the classical case (i.e. $\hbar \rightarrow 0$).
 
\subsection{CASE II (Saddle-node bifurcation)} The saddle-node bifurcation corresponds to the Hamiltonian given by
\begin{eqnarray}
H={\frac 1 2}(p^2+\Delta p^2)-
[(\epsilon x + {\frac {\beta x^3} 3} 
+ {\frac{\Delta x^2} 2}(2\beta x)]\sum_n\delta(t-nT).\nonumber 
\end{eqnarray}
Substituting for $\Delta x^2$ and $\Delta p^2$ respectively we get
\begin{eqnarray}
H = [{\frac {p^2} 2} + {\frac \hbar 2}({\frac 1 {4\phi}} + 4\pi^2 
\phi)]  - [(\epsilon x + {\frac {\beta x^3} 3})
+ \hbar \phi\beta x ] \sum_n\delta(t-nT).
\label{eq9}
\end{eqnarray}
The corresponding difference equations are 
\begin{eqnarray}
X_{n+1}=X_n + P_n,
\label{eq10}
\end{eqnarray}
\begin{eqnarray}
P_{n+1}=P_n + T[\epsilon + \beta X_n^2 + {\frac{T\beta \hbar \Phi _n}  2}],
\label{eq11}
\end{eqnarray}
\begin{eqnarray}
\Phi _{n+1} = \Phi _nF_n, F_n = {\frac 1 {\Phi _n^2}} + (\Pi _n+1)^2
\label{eq12}
\end{eqnarray}
and
\begin{eqnarray}
\Pi _{n+1} = [\Pi _n + \Pi _n^2 + {\frac 1 {\Phi _n^2}}]{\frac 1 F_n} + 
2T\Delta _{n+1},
\label{eq13}
\end{eqnarray}
where
$\Delta _{n+1} = \beta X_{n+1}$
(the symbols have their usual meanings).
In the classical case ($\hbar \rightarrow 0$) these correspond to the existence of a stable
fixed point $X=-\sqrt(-{\frac\epsilon \beta}), P=0$ for $\epsilon <0,\beta>0$ 
while for $\epsilon>0,\beta>0$ no fixed point exists. This leads to chaos via 
intermittency. 

The phase space is divided into two 
regions, one bounded by $X=\pm X_0, P=\pm P_0$ ($X_0$ and $P_0$ have suitably chosen magnitudes 
$S$ and $S'$ respectively)
and called the laminar region, while the remaining region can be said to belong to the chaotic phase. Thus the
journey in phase space will consist of regions of laminar behaviour interrupted by alternate 
regions of chaotic phase, characteristic of intermittency. The boundedness of the phase space ensures reentry of the phase points 
in the laminar region after a turbulent state and provided the number of iterations is large 
enough and the initial point suitably chosen, the reentry will be a uniform or 
white reentry. The average staying time in the laminar phase is calculated by 
assuming a uniform reentry which can also be simulated in the numerical calculation 
by starting with a uniformly distributed choice of initial points in the region. We have extended our calculations also to a non-uniform reentry. 
The numerical results remain qualitatively the same.
        
\section{Numerical Results and Discussions}

The signatures of quantum effects in classical intermittency through 
bifurcations, albeit in the semi-quantal domain, were investigated by 
(i) considering the phase space plots
and (ii) probing into other hallmarks of chaos (given later) relevant 
to two maps exhibiting saddle-node and inverted pitchfork bifurcations 
respectively. The exit of the phase space points starting from the
neighbourhood of stable fixed point becomes, as is evident from the 
Figs. 1-3, progressively more and more rapid with increase of $\hbar$
in both the maps, thus supporting our conjectures about the quantum
enhancement of chaos. The quantum tunneling effects are demonstrated
by the plots (Fig. 2), where the system is a single particle periodically 
kicked by a 
potential which is a double well with a single hump in 
between, with the initial point chosen to be close to any one of the 
stable minima. The other plots given subsequently (Figs. 4-11) corroborate our view in this regard. 

The relevant features of
chaos, namely (i) the average staying time in the laminar phase, (ii) the Lyapunov
exponent, (iii) the energy diffusion were determined for both the above maps. Since
we have endeavoured to probe into the situation on either side of the 
bifurcation point, the magnitudes of the parameters $\epsilon, \beta$ were 
chosen accordingly. For the inverted pitchfork bifurcation the parameters 
chosen were (i) $\epsilon >0, \beta >0$, (ii) $\epsilon <0, \beta >0$ and
(iii) $\epsilon >0, \beta <0$ respectively. The first case 
corresponds to the existence of an unstable fixed point at $X = 0, P = 0$, while
the case (ii) corresponds to the presence of three fixed points of which the 
pair at $X = \pm \sqrt(-{\frac \epsilon \beta}), P = 0$ are unstable and the 
one at $X =0, P = 0$ is stable (clearly here $\epsilon = 0$ is the local 
bifurcation parameter). The case (iii) corresponds to supercritical bifurcation
and we have included
this case as well to show the effect of two stable fixed points at 
$X = \pm \sqrt(-{\frac \epsilon \beta}), P = 0$ on the relevant parameters both
in the presence and the absence of quantum effects. 

The numerical calculations have been performed in the following way:
(a) In the calculation of the average staying time in the laminar phase an
ensemble of initial points (assuming both uniform as well as non-uniform distribution) surrounding the stable fixed point was chosen and
for each of these the number of iterations required for exit out of the laminar region ($S = S' =$ 0.1 here) was 
calculated. The average was calculated for each value of $\hbar$, 
(b) for calculation of the average Lyapunov exponent $\lambda$, a pair of 
closely lying initial points (with separation, say, $d_0$) was chosen near 
the stable fixed point. After every two iterations the distance $d$ between the two
points was calculated and the expression $\lambda_1 = {\frac 1 2}\ln{\frac d {d_0}}$ determined numerically. The distance was normalized to the initial value
and the process repeated till the evolved points left the laminar region. The
resultant average Lyapunov exponent $\lambda$ over the entire laminar region 
was then computed, (c) lastly, for the estimate of the energy diffusion, average
value of $p^2$ was calculated for a choice of the initial points lying close to the stable fixed point for each value of $\hbar$. In all these cases, the 
calculations were repeated for different choices of the parameters and initial 
points respectively. Since the results obtained were similar, only a typical 
set has been displayed in the manuscript.

The influence of the stable fixed points
is evidently manifested, at least for low quantum effects by (i) the increase of 
average staying time in the laminar phase, (ii) the decrease of Lyapunov exponent, (iii) the resistance to the energy diffusion 
compared to that in the absence of the fixed point. For 
example, the parameters (a) $\epsilon <0, \beta >0$ and (b) $\epsilon >0, \beta <0$ correspond to the existence,
 while the parameters
(c) $\epsilon >0, \beta >0$ pertain to the absence of the stable fixed points in the inverted pitchfork bifurcations. For the saddle-node bifurcation
we have investigated the case $\epsilon <0, \beta >0$ as well as the case 
$\epsilon >0, \beta >0$. The first one corresponds to a stable fixed point at
$X = -\sqrt(-{\frac \epsilon \beta}), P = 0$ while there is no fixed point on
the other side of the bifurcation point $\epsilon >0, \beta >0$ in the second 
case (The case $\epsilon>0,\beta<0$ is uninteresting in this case). A range of 
possible values of $\epsilon, \beta$ was covered but since 
the results were found out to be qualitatively similar, the results for a 
typical choice of parameters have been displayed here.

It should be noted that although the effects of quantum fluctuations on
classical dynamics have been investigated in some standard Hamiltonian maps
\cite{hu2} no such 
attempt has been made earlier in problems involving chaos via routes of 
intermittency through bifurcations. The semi-quantal formalism gives rise to a 
set of four coupled equations involving $x$, $p$, $(\Delta x)^2$,
$(\Delta p)^2$. The effect of the quantum fluctuations $(\Delta x)^2$, $(\Delta p)^2$ on the evolution of the wavepacket leads for a non-compact phase space 
relevant to our problem to an enhancement of chaos. Analogous conclusions have
been reached in some earlier works\cite{won,hu2,jac}. 

In both the maps we have dealt with the situations corresponding to the presence
of unstable fixed points (or no fixed point in the case of saddle-node 
bifurcations) as well as stable fixed points lying on the opposite sides of the
bifurcation points respectively. The variations of the relevant parameters with the increase
of quantum effects are shown graphically. The numerical calculations yield the 
following features corresponding to both the maps.

If we start with initial points lying close to stable fixed points (in the basin
of attraction of the fixed point) (which are (0,0) for $\epsilon <0, \beta >0$,  and 
$(\pm\sqrt(-{\frac \epsilon \beta}),0)$ for $\epsilon >0, \beta <0$ in the case of
the first map and $(-\sqrt(-{\frac \epsilon \beta}),0)$ for the second map respectively), the 
average staying times in the laminar phase will be, for low $\hbar$, greater
than those in the case corresponding to the absence of any stable fixed point 
($\epsilon >0, \beta >0$ or $\epsilon < 0, \beta < 0$ in both the maps). This is quite expected as a 
consequence of the presence of the stable fixed point in an almost classical
domain. Contrary to classical dynamics, where the initial points close to the stable fixed point remain always in its neighbourhood, there is a
wandering of the points away from the stable fixed point for large quantum effects, i.e., for large values of $\hbar$.  This is also amply manifested in the
phase space plots shown (Figs. 1-3). 

The classical effects are seen to be offset by the introduction of
quantum fluctuations which decrease the average staying times in the laminar 
phase leading to 
practically identical values for large values of $\hbar$ (Figs. 4 and 5). 
In respect of the average Lyapunov exponent $\lambda$, the observations are similar except that 
$\lambda$ increases with the increase of $\hbar$ (Figs. 6 and 7) and for 
low $\hbar$ it is smaller in the presence of a stable fixed point. 
Regarding the energy 
diffusion, the presence of the stable fixed point decreases the rate of 
diffusion compared to the case of no-stable fixed point for low value of 
$\hbar$. Moreover the energy
remains bounded within certain limits leading to a certain kind of localisation
in the case of the initial points chosen close to the stable fixed points at least for
low quantum effects (Figs. 8 and 10). For higher quantum effects the energy diffusion becomes 
much more rapid (Figs. 9 and 11). However, for large quantum effect, contrary to
the classical case, no simple scaling law\cite{lah} regarding the variation with 
$\epsilon$ has been seen to exist. Though we have confined ourselves to the 
laminar region, the situation is expected to be similar for large $\hbar$ in the case of the 
turbulent region which lies outside the gate bounding the laminar phase.
A different choice of initial $X$ and $P$, for example, both lying in the region
of classically unstable fixed points has been seen to display a similar behaviour. 

The enhancement of stochasticity (reflected in the decrease of the average 
staying time in the laminar phase, increase of the average Lyapunov exponent as
well as the rate of energy diffusion with the onset of quantum effects) can be 
qualitatively explained by condsidering predominantly the non-compactness of 
the phase space and also partly by taking into account the broadening of the 
wavepacket as well as the quantum mechanical tunneling. The significance of 
the typical quantum mechanical features is revealed in the graphical plots 
where the preponderance of the quantum mechanical effects is clearly visible 
even when the initial points are chosen to lie in the neighbourhood of the 
classical stable fixed points. Similar quantum mechanical enhancement of chaos 
has also been observed if the initial points are chosen to lie near the 
classically unstable fixed points (although not shown in the plots displayed 
here). It is, therefore, clear from our results utilizing the squeezed coherent
state formalism that the quantum tunneling and wavepacket spreading 
effects combined with the non-compactness of the underlying phase space 
dominate over the chaos-suppressing interference effects\cite{gut} for the 
type of bifurcations investigated. Similar results have also been obtained by 
the earlier workers\cite{won,hu2}, albeit in other contexts. 

The squeezed coherent state formalism thus enlightens us about the effects of
quantum fluctuations on the problems of intermittency involving bifurcations of
the saddle node and the inverted pitchfork types respectively. This semiquantal
formalism admits of applications to other problems of classical chaos.

\section*{Acknowledgements}
We express our indebtedness to Profs. A. Lahiri, G. Ghosh, P. Mukherjee and M. Acharyya for help and suggestions.

\newpage

\begin{center}{\bf Figure Captions}\end{center}

{\bf Fig-1.} Phase space plots for the inverted pitchfork bifurcation 
($\epsilon=-10^{-4}$, $\beta=1.0$, $\phi = 0.5$ (initial value), $\pi = 0.0$ (initial value) and $T=1.0$) with the initial point close to the stable fixed point
$X_0 = 0, P_0 = 0$ for $\hbar$ = 0 (left figure), = 10$^{-6}$ (middle figure), 
= 1.0 (right figure). The enhancement of chaos with the increase of quantum 
effects is reflected in the progressively rapid exit of the phase space points
with the increase of $\hbar$.

{\bf Fig-2.} Phase space plots for the inverted pitchfork bifurcation 
($\epsilon=10^{-4}$, $\beta=-1.0$, $\phi = 0.5$ (initial value), $\pi = 0.0$ (initial value) and $T=1.0$) with the initial point close to the stable fixed point
$X_0 = \pm 0.01, P_0 = 0$ for $\hbar$ = 0 (left figure), = 10$^{-6}$ (middle 
figure), = 1.0 (right figure). The enhancement of chaos with the increase of 
quantum effects is reflected in the progressively rapid exit of the phase space 
points with the increase of $\hbar$.

{\bf Fig-3.} Phase space plots for the saddle node bifurcation 
($\epsilon=-10^{-4}$, $\beta=1.0$, $\phi = 0.5$ (initial value), $\pi = 0.0$ (initial value) and $T=1.0$) with the initial point close to the stable fixed point
$X_0 = -0.01, P_0 = 0$ for $\hbar$ = 0 (left figure), = 10$^{-6}$ (middle 
figure), = 1.0 (right figure). The enhancement of chaos with the increase of 
quantum effects is reflected in the progressively rapid exit of the phase space
points with the increase of $\hbar$.

{\bf Fig-4.} Plot of average staying time against $\log_{10}(\hbar)$ for the maps 
exhibiting inverted pitchfork bifurcation. The three graphs correspond to the 
cases a) $\epsilon>0,\beta>0$, b) $\epsilon<0,\beta>0$ and c) $\epsilon>0,
\beta<0$ respectively.  The parameters 
chosen are $\epsilon=10^{-4}$, $\beta=1.0$, $\phi = 1.5$ (initial value), $\pi = 0.0$ (initial value) and $T=1.0$ respectively.
The laminar region is given by $X_0 = P_0 = 0.1$.
The phase space is chosen as a square of size 1.0.
The decrease of average staying time with the 
increase of quantum effect is reflected in the figure. The results shown here
are for a uniform distribution of initial points although they were seen to be
similar when this distribution is non-uniform. 

{\bf Fig-5.} Plot of average staying time against $\log_{10}(\hbar)$ for the maps exhibiting 
saddle-node bifurcation. The two graphs correspond to the cases a) $\epsilon>0,\beta>0$ and b)
        $\epsilon<0,\beta>0$ respectively.
The parameters are identical with those in the first case.
The laminar region and the phase space are chosen as before.The results shown here
are for a uniform distribution of initial points although they were seen to be 
similar when this distribution is non-uniform.

{\bf Fig-6.} Plot of $\log_{10}$(average Lyapunov exponent) against $\log_{10}(\hbar)$ for the maps 
exhibiting inverted 
pitchfork bifurcation.The three graphs correspond to the cases a) $\epsilon>0,\beta>0$,
b) $\epsilon<0,\beta>0$ and c) $\epsilon>0,\beta<0$ respectively. The initial points are
        chosen to lie in the neighbourhood of the stable fixed point for the last two cases.
        The parameters chosen are the same as in the earlier cases except that 
the initial value of $\phi$ is 0.5. 
$d_0$ (see text) is chosen close to $10^{-9}$.

{\bf Fig-7.} Plot of $\log_{10}$(average Lyapunov exponent) against $\log_{10}(\hbar)$ for the maps exhibiting saddle-node
bifurcation.The two graphs correspond to the cases a) $\epsilon>0,\beta>0$,
b) $\epsilon<0,\beta>0$ respectively. The initial points are chosen to lie in 
the neighbourhood of the stable fixed point for the second case. The parameters chosen are the same as in Fig. 6. 

{\bf Fig-8.} Plot of $\log_{10}$(energy) against the iteration number {\it a} for the maps exhibiting inverted pitchfork bifurcation in the cases 
a) $\epsilon > 0, \beta > 0$,  
b) $\epsilon < 0, \beta > 0$ and       
c) $\epsilon > 0, \beta < 0$ respectively. These plots refer to low quantum effect ($\hbar = 10^{-10}$). 
The values of the parameters are chosen as in Fig. 7.
The initial points are chosen to lie in the neighbourhood of stable fixed 
points for the last two cases.

{\bf Fig-9.} Plot of $\log_{10}$(energy) against the iteration number {\it a} for the maps exhibiting inverted pitchfork bifurcation in the cases 
a) $\epsilon > 0, \beta > 0$ (the solid line),  
b) $\epsilon < 0, \beta > 0$ (the long-dashed line) and       
c) $\epsilon > 0, \beta < 0$ (the short-dashed line) respectively. These plots 
refer to large quantum effect ($\hbar = 10^{-2}$). 
The values of the parameters are chosen as in Fig. 7.
The initial points are chosen to lie in the neighbourhood of stable fixed points for the last
two cases. The first two cases (the solid line and the long-dashed line)
give almost identical results for this value of $\hbar$ and hence are indistinguishable
in this plot.

{\bf Fig-10.} Plot of $\log_{10}$(energy) against the iteration number {\it a} for the maps 
exhibiting saddle-node bifurcation in the cases 
a) $\epsilon > 0, \beta > 0$ and b) $\epsilon < 0, \beta > 0$     
respectively. These plots 
refer to low quantum effect ($\hbar = 10^{-10}$). The values of the parameters are chosen as in Fig. 7. 
The initial points are chosen to lie in the neighbourhood of the stable fixed point for 
the second case.

{\bf Fig-11.} Plot of $\log_{10}$(energy) against the iteration number {\it a} for the maps 
exhibiting saddle-node bifurcation in the cases 
a) $\epsilon > 0, \beta > 0$ and
b) $\epsilon < 0, \beta > 0$     
respectively. These plots 
refer to large quantum effect ($\hbar = 10^{-3}$). The values of the parameters are chosen as in Fig. 7.
The initial points are chosen to lie in the neighbourhood of the stable fixed point for 
the second case.
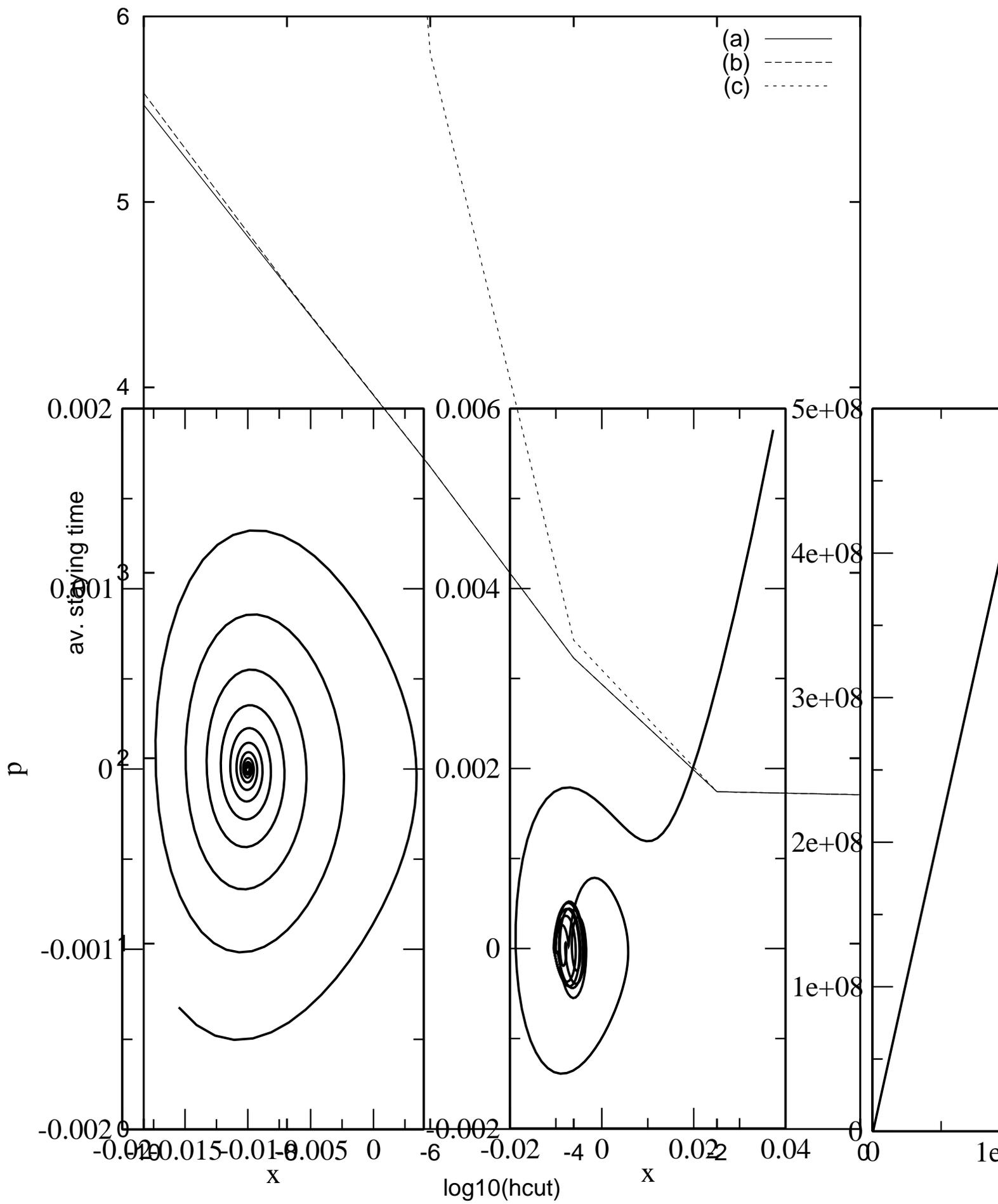
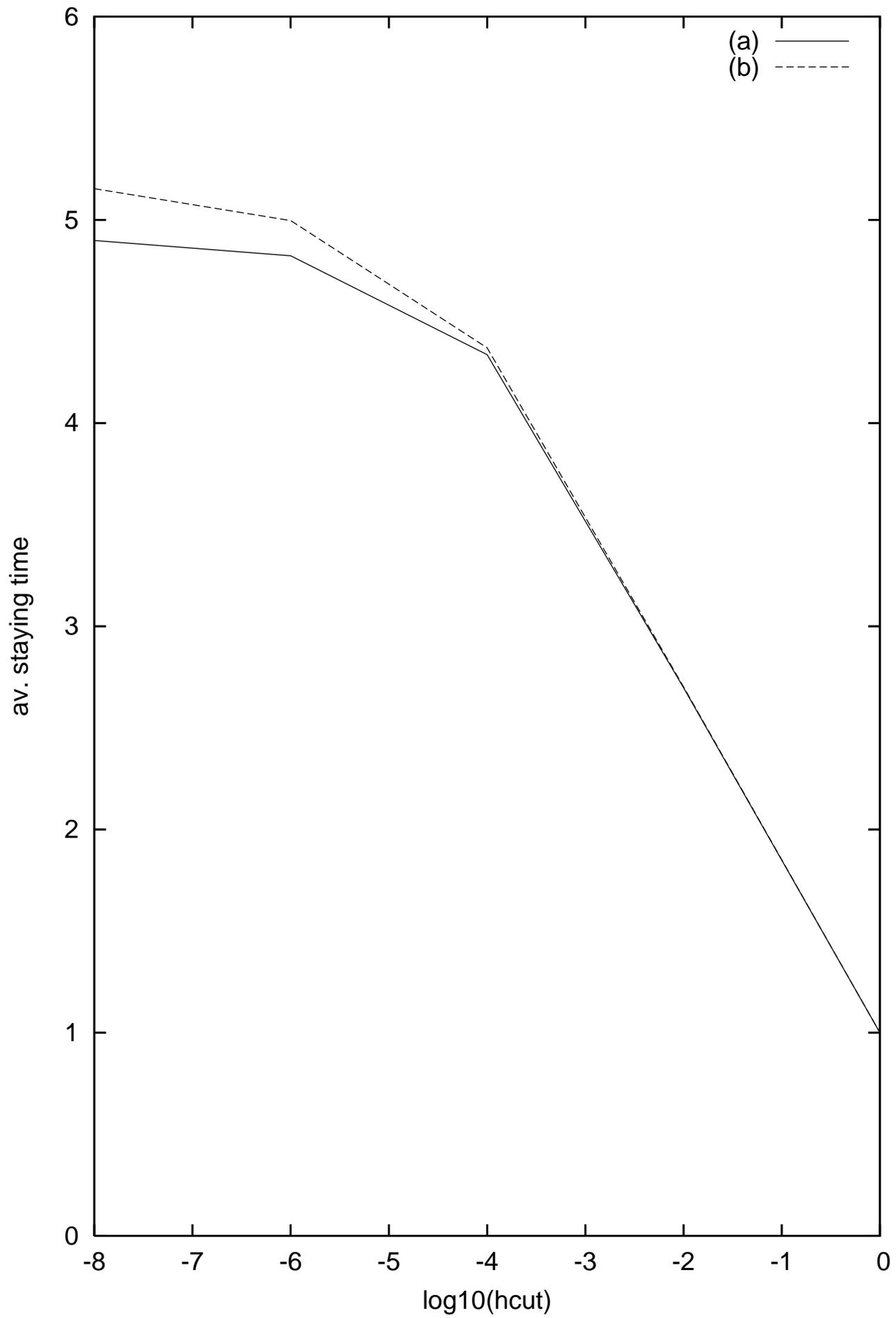
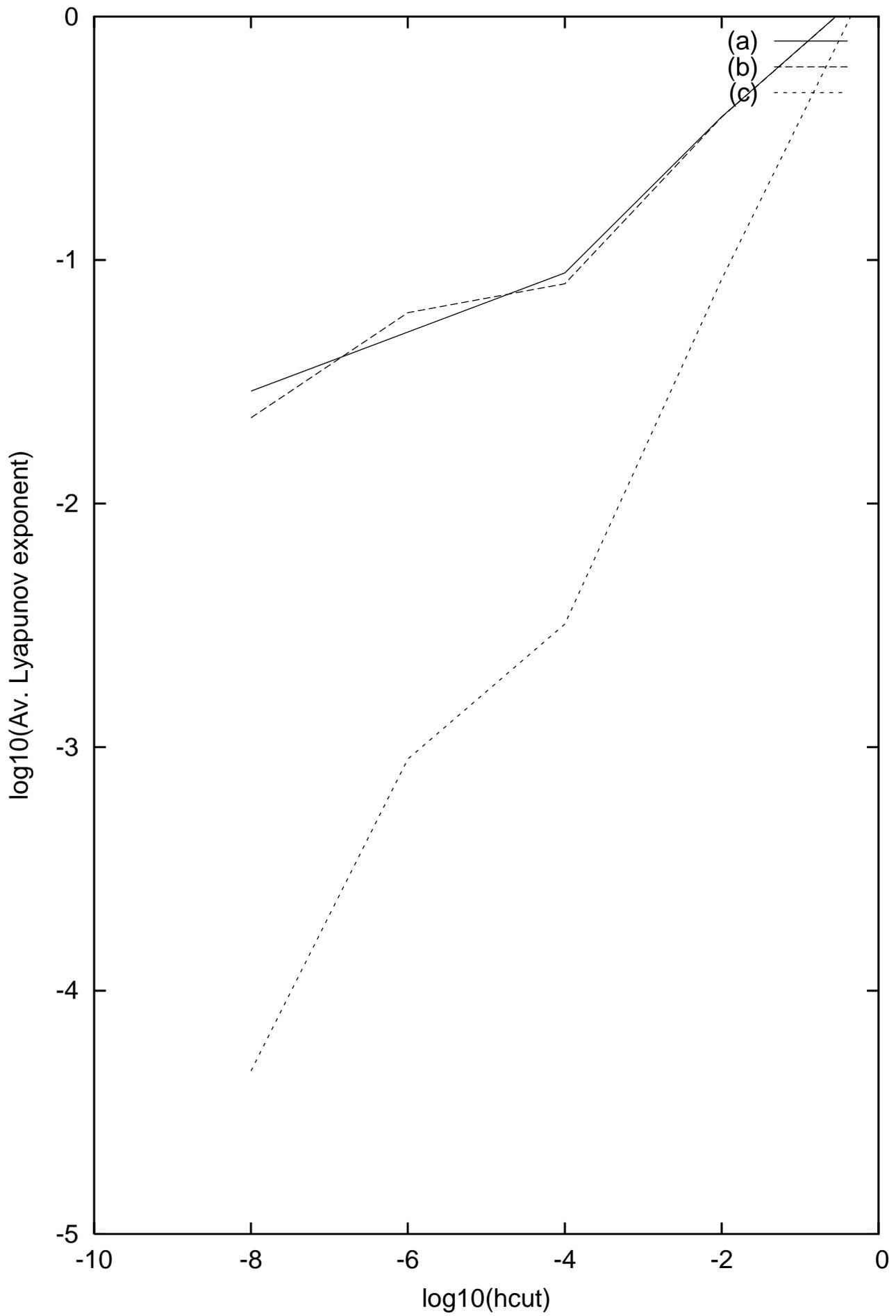
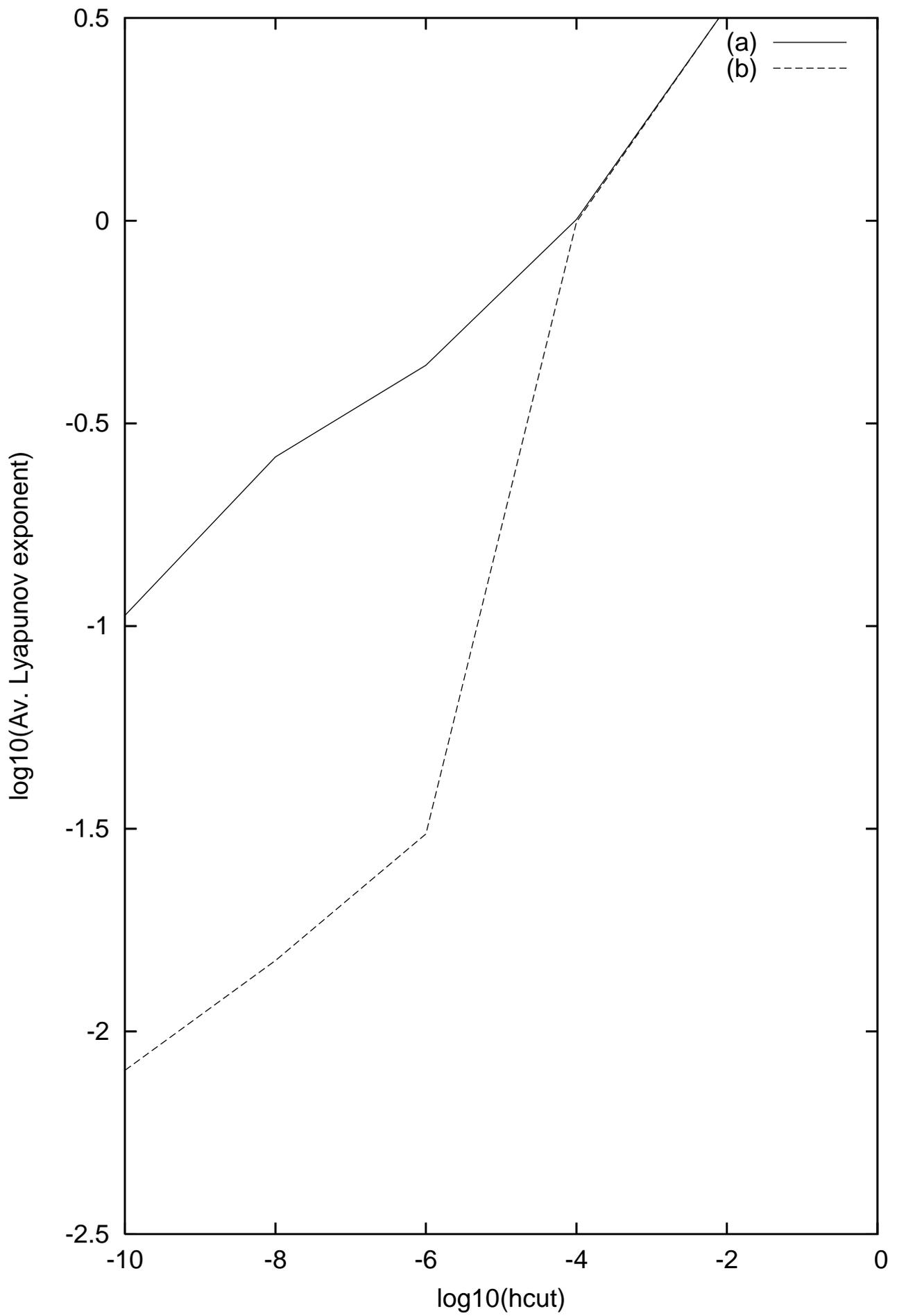
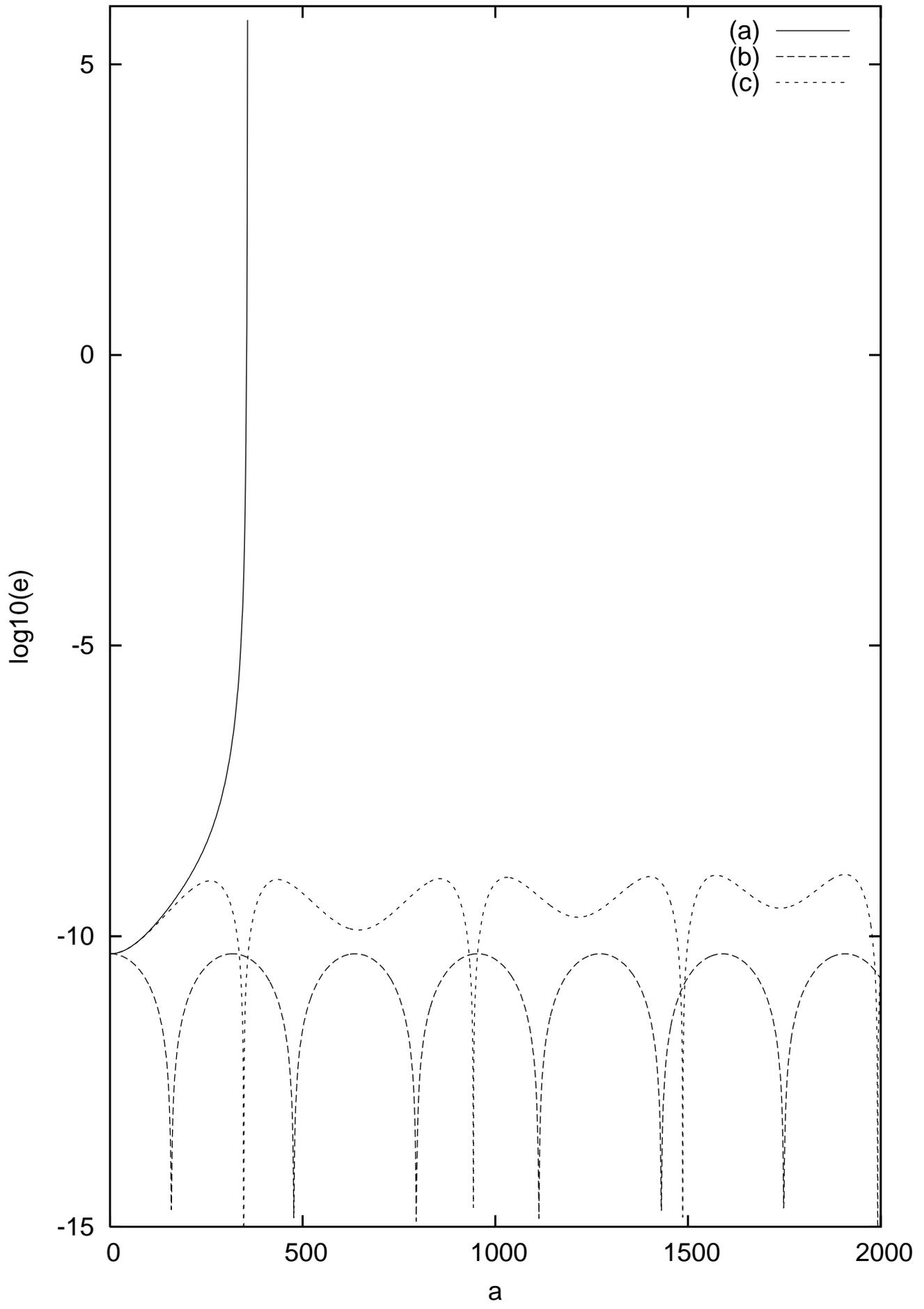
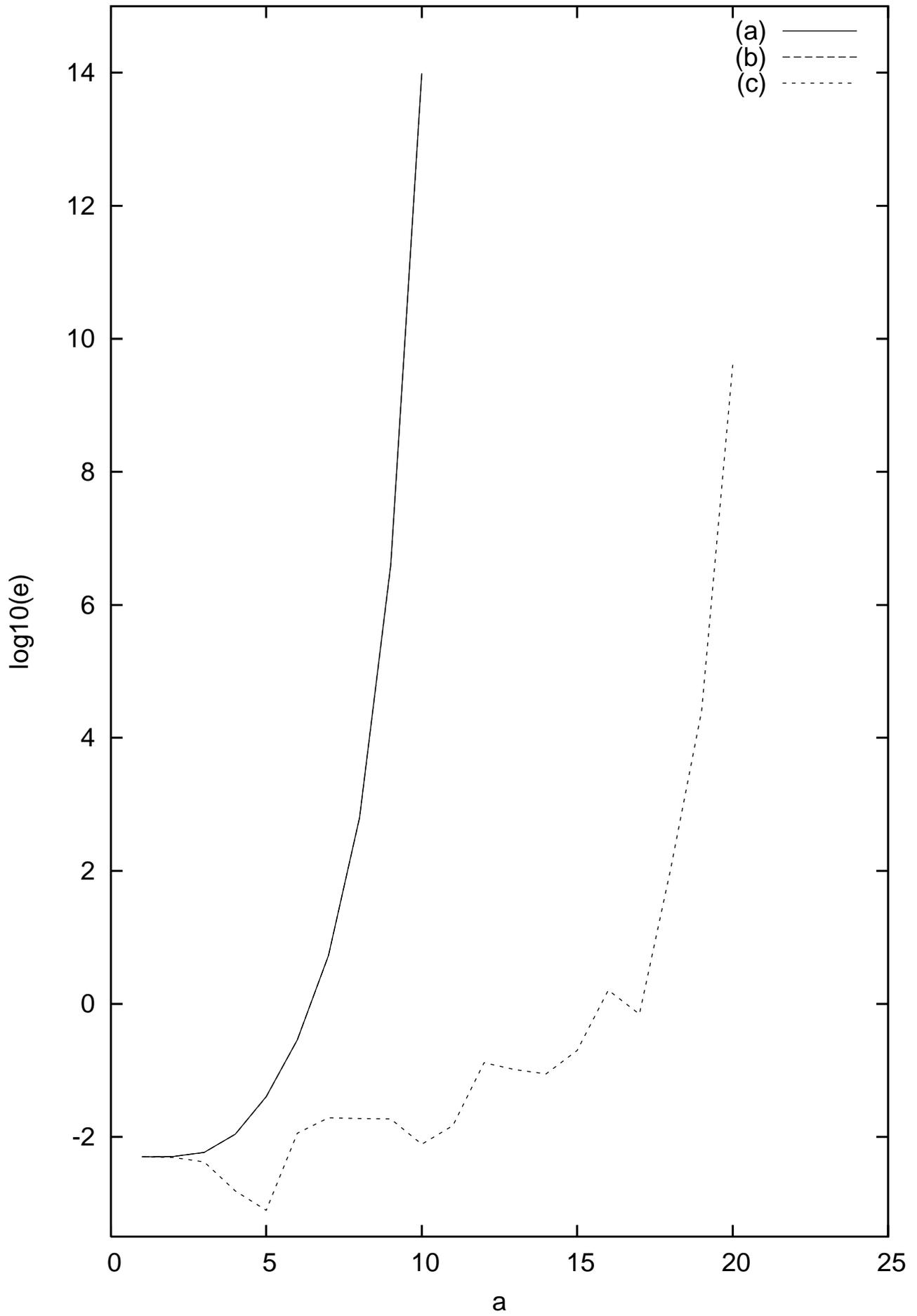
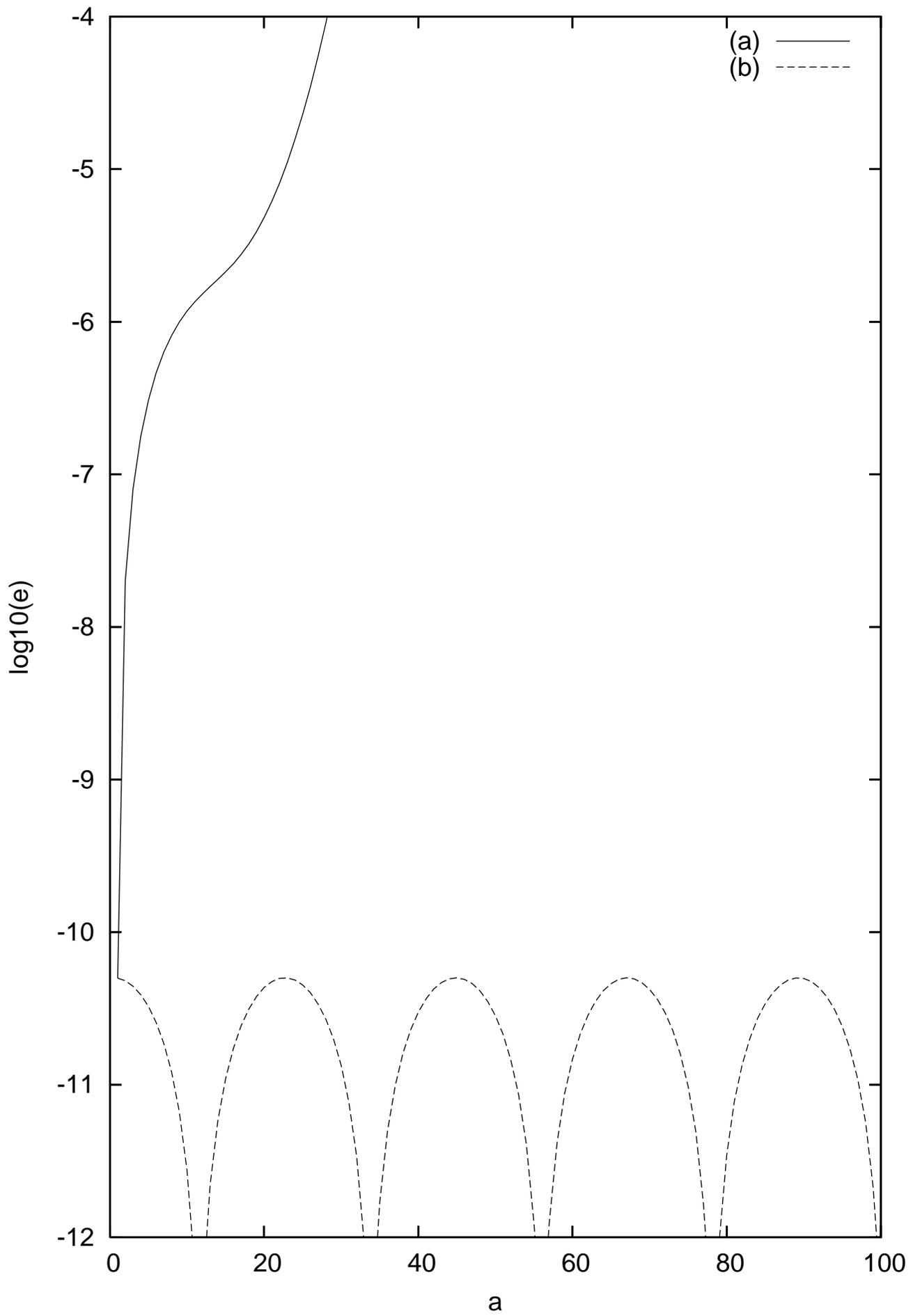
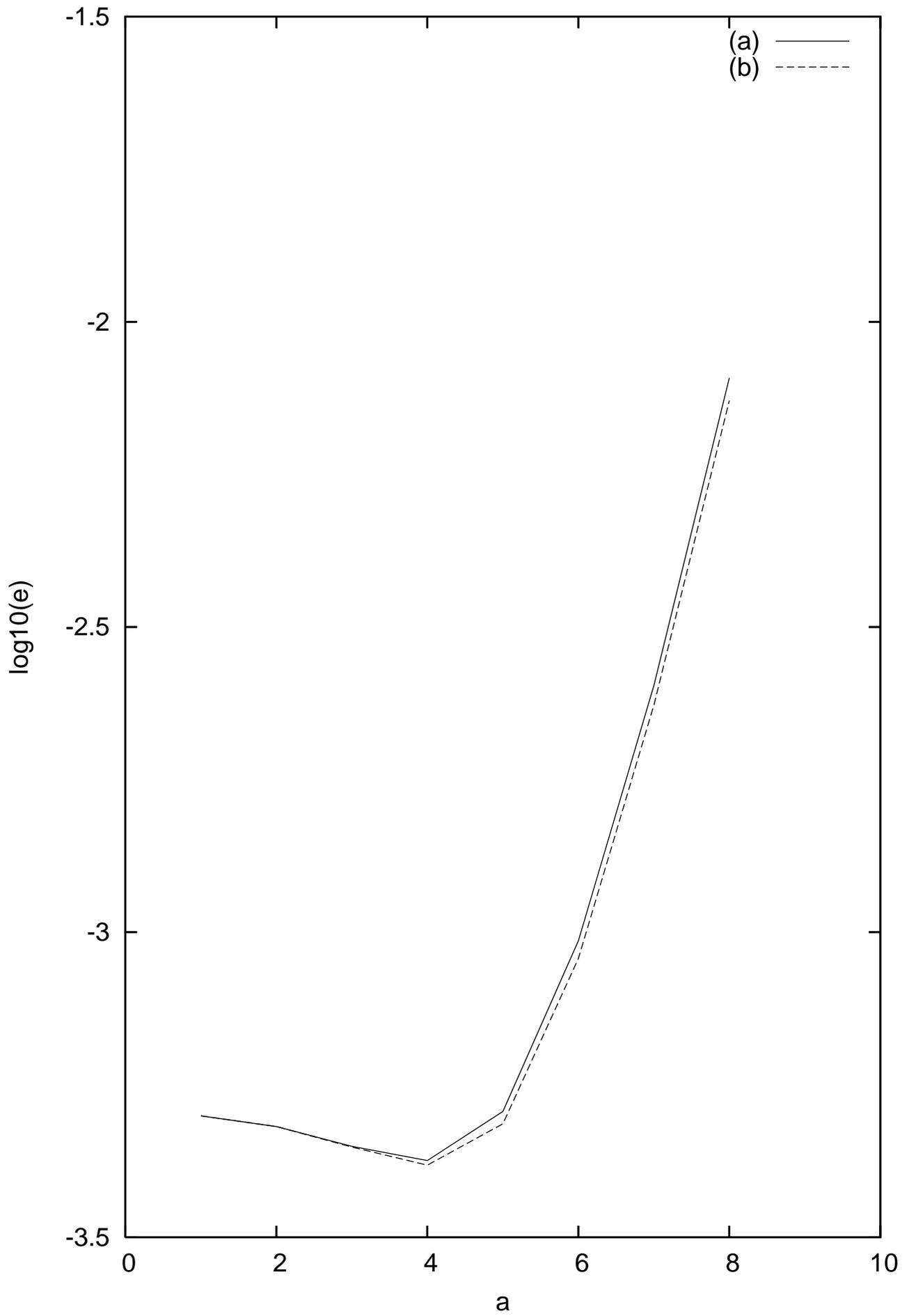
\end{document}